\documentstyle[11pt,paspconf,epsf,twoside]{article}
\begin{document}
\title{The NOG sample:\\ 
selection of the sample and identification of galaxy systems}  
\author{Giuliano Giuricin, Christian Marinoni, Lorenzo Ceriani, Armando 
Pisani} 
\affil{Dept. of Astronomy, Trieste Univ. and SISSA, Trieste, Italy}

\begin{abstract}
In order to map the galaxy density field in the local
universe, we select the  Nearby Optical  Galaxy  (NOG) sample, which is
a volume-limited ($cz\leq$6000 km/s) and  magnitude--limited ($B\leq$14
mag) sample of 7076 optical  galaxies which covers 2/3 (8.29 sr) of the 
sky ($|b|>20^{\circ}$) and has a good completeness in redshift (98\%). 

In order to trace the galaxy density field on small scales, we identify 
the NOG galaxy systems by means of both the hierarchical and the 
percolation {\it friends of friends} methods.

The NOG provides high resolution in both spatial sampling of the nearby
universe and morphological galaxy classification. The NOG is meant to be
the first step towards the construction of a statistically well-controlled
galaxy sample with homogenized photometric data covering most of the
celestial sphere.
\end{abstract}

\section{The selection of the NOG} 

Relying, in general, on photometric data tabulated in LEDA (Paturel et al. 
1997), we select the Nearby Optical Galaxies (NOG) sample, a sample of
7076 galaxies which lie at $|b|>20^{\circ}$, have recession velocities (in
the Local Group frame) $cz\leq$6000 km/s, and have corrected total blue
magnitudes B$\leq$14 mag. We have incorporated numerous new redshifts
given by the Optical Redshift Survey (Santiago et al. 1995), the Updated
Zwicky Catalog (Falco et al.  1999), and the newly completed PSCz survey
(for which data have been kindly provided to us by William Saunders). 

We select the sample attempting to tighten the selection criteria already 
used in the ORS survey, specifically, by adopting a complementary approach
to the construction of an all-sky optical galaxy sample.
Specifically, we use, as photometric selection parameter, the  total blue 
magnitudes, homogeneously
transformed into the standard system of the RC3 catalogue and corrected 
for Galactic extinction, internal extinction and K-dimming. This is meant
to  minimize systematic selection effects as a function of direction in the 
sky and,thus, provide a largely unbiased view of the galaxy distribution. We 
limit the sample to the depth of 6000 km/s in order to reduce  
the incompleteness in redshift and have a dense sampling of the
galaxy density field. From the analysis of the behaviour of galaxy counts
versus magnitude, we find that the NOG is intrinsically complete
down to its limiting magnitude B=14 mag. Moreover, it has a good 
completeness in redshift ($\sim$98\%). 

Fig. 1 shows the NOG galaxies and the IRAS 1.2 Jy sample
(Fisher et al. 1995) limited to 6000 km/s, using Flamsteed projections in
Galactic coordinates. The region devoid of galaxies is the zone of
avoidance. Similar, prominent structures stand out in these plots such as the
densest part of the Local Supercluster at $l=300^{\circ}-315$,
$b=30^{\circ}-75^{\circ}$ with the Virgo cluster at $l=284^{\circ}$,
$b=75^{\circ}$, the Hydra-Centaurus complex (around $b=20^{\circ}$,
$l=260^{\circ}-310^{\circ}$), together with the contiguous
Telescopium-Pavo-Indus supercluster (from $b=-20^{\circ}$, $l=330^{\circ}$
to $b=-60^{\circ}$, $l=30^{\circ}$ and the Perseus-Pisces supercluster 
($l=110^{\circ}-150^{\circ}$). There are also pronounced voids such as 
the Local Void, which covers a large part of the sky between 
$l=0^{\circ}$ and $l=80^{\circ}$, the Orion-Taurus void 
($l=150^{\circ}-190^{\circ}$, $b\sim-30^{\circ}$), and the Eridanus void 
(around $l=270^{\circ}$, $b=-65^{\circ}$).

\section{The Identification of Galaxy Systems}

Since we are interested in describing the galaxy density field also on
small scales, we identify galaxy systems in the NOG by means of the
hierarchical method (according to the precepts given by Gourgoulhon et al.
1992) and percolation {\it friends of friends} method (with the distance
and velocity link parameters chosen as in Ramella et al. 1997).

We obtain two homogeneous catalogs of loose groups which turn out to be
substantially consistent. Containing about 500 groups with at least three
members, they are among the largest catalogs of groups presently available
in the literature. Most of the NOG galaxies ($\sim$60\%) are found to be
members of galaxy pairs ($\sim$580 for a total of $\sim$15\% of the
galaxies) or groups with at least three members ($\sim$500 groups
comprising $\sim$45\% of the galaxies). About 40\% of the galaxies are
left ungrouped (field galaxies).

\begin{figure}
\plotfiddle{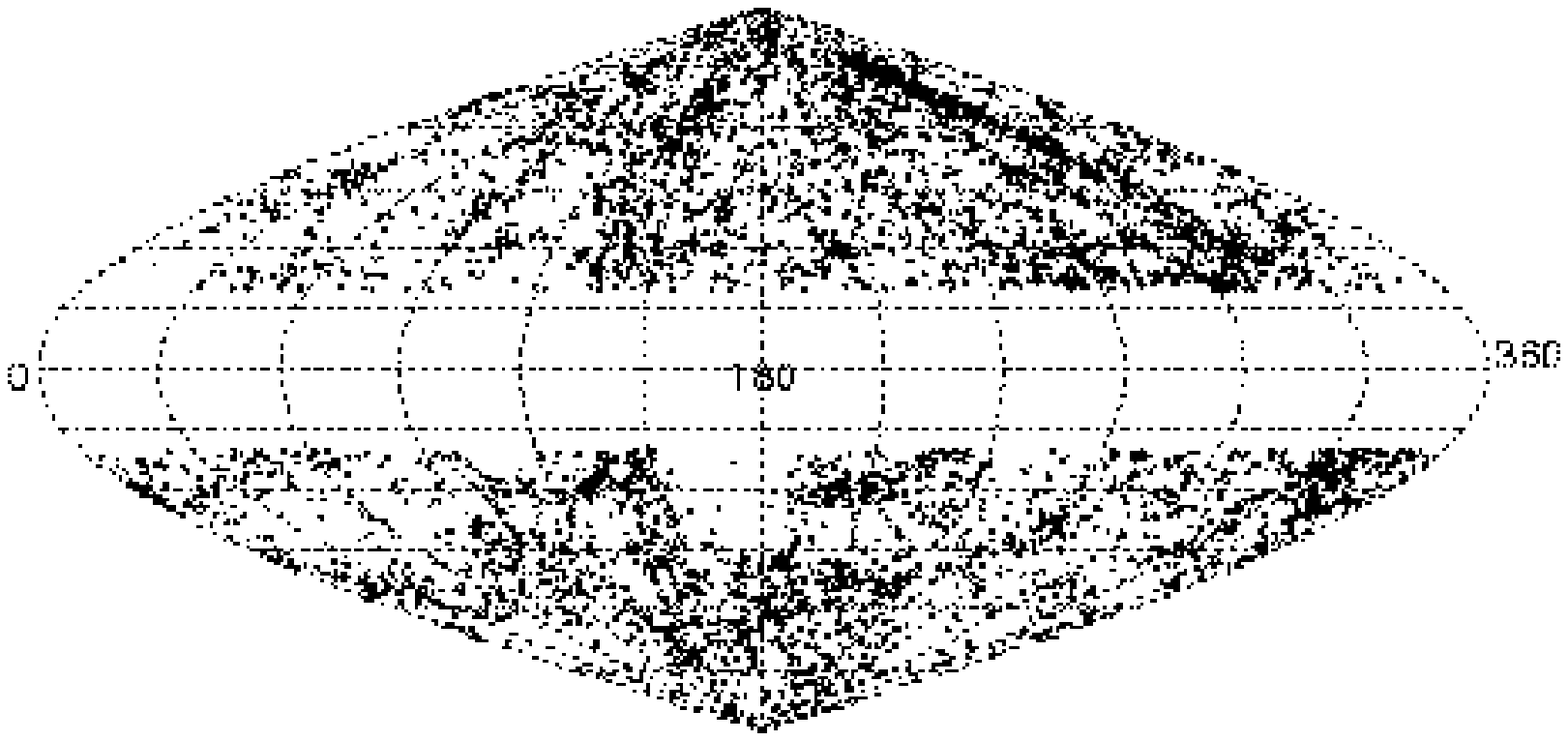}{7.5cm}{0}{80}{80}{-285}{-200}
\plotfiddle{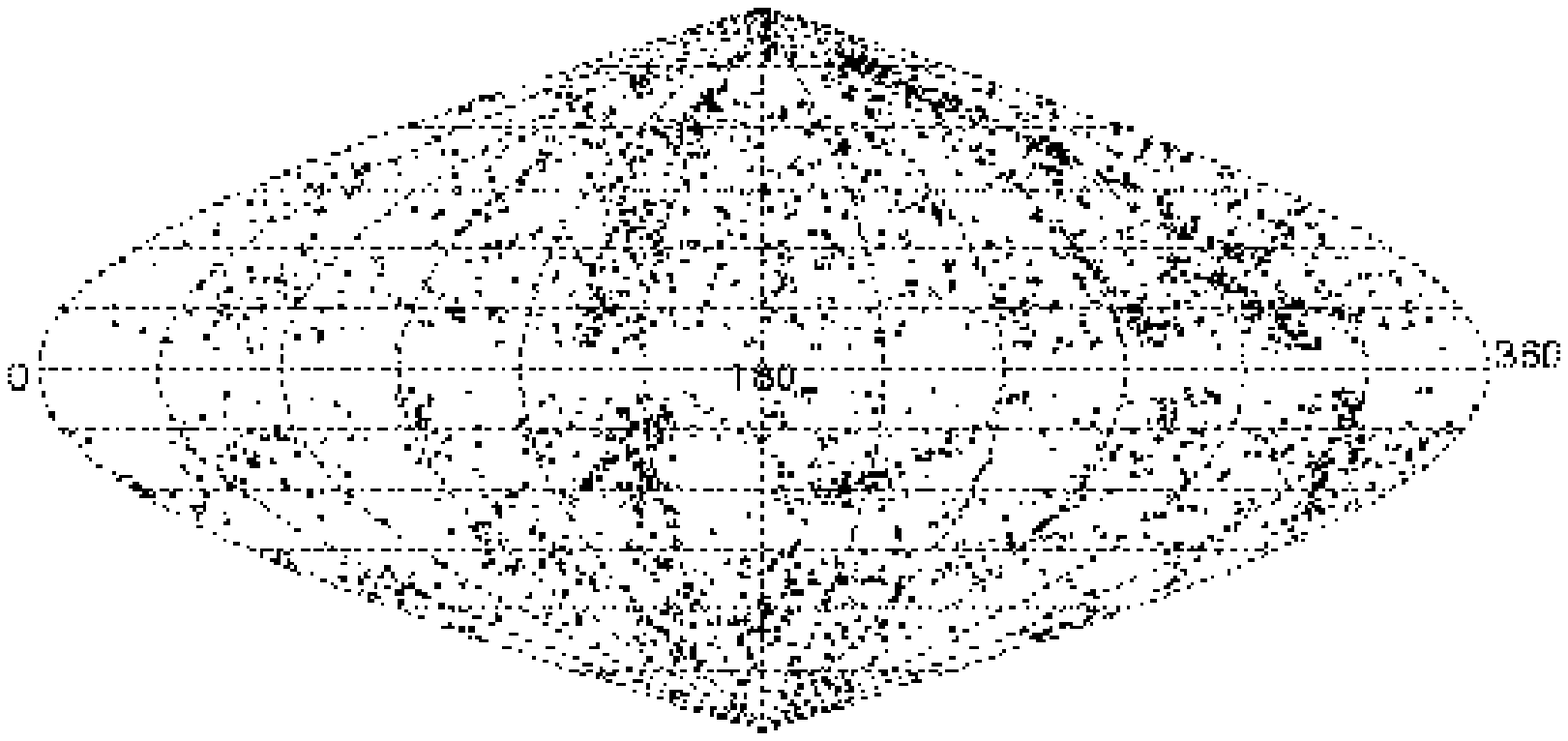}{7.5cm}{0}{80}{80}{-285}{-200} 
\caption{{\em Upper:} The NOG sample is shown in Flamsteed projection on 
the celestial sphere using Galactic coordinates. {\em Lower:} 
The IRAS 1.2 Jy sample by Fisher et al. 1995 limited to 
6000 km $s^{-1}$ is shown in the same projection. In spite of the smoothing 
caused by projection over distance, prominent structures and voids are still 
visible.}
\end{figure}

\section{Conclusions}

The NOG groups will be used to remove non-linearities in the peculiar
velocity field on small scales. To correct the redshift-distances of field
galaxies and groups on large scales, we shall apply models of the peculiar
velocity field, following the approach described in Marinoni et al.
(1998). We shall use the locations of individual galaxies and groups
calculated in real-distance space to calculate the selection function of
the NOG sample (following the approach described in Marinoni et al. 1999)
and to reconstructed the galaxy density field (see the paper by Marinoni
et al. in the same volume). 
 
Though being limited to a depth of 6000 km/s, the NOG covers interesting
regions of galaxy and mass overdensities of the local universe, such as
the "Great Attractor" region and the Perseus-Pisces supercluster. 
Compared to previous all-sky optical and IRAS galaxy samples (e.g. the
Optical Redshift Survey by Santiago et al. 1995, the IRAS 1.2 Jy by Fisher
et al.  1995, the PSCz by Saunders et al. 1999), the NOG provides a denser
sampling of the galaxy density field in the nearby universe. NOG contains
11\% more galaxies than the ORS limited to 6000 km/s and 35\% more
galaxies than the PSCz limited to the same depth, although it covers about
3/4 of the solid angle covered by the PSCz. Besides, NOG delineates
overdensity regions with a greater contrast than IRAS samples do (see Fig.
1). 

Given its large sky coverage, its high-density sampling, and the 
identification of galaxy systems, the NOG is well suited to mapping the 
cosmography of the nearby universe, studying the clustering properties 
and tracing the optical galaxy density 
field (also on small scales) to be compared with the IRAS galaxy density 
field. Local galaxy density parameters derived from the NOG are meant to 
be used in statistical investigations of environmental effects on nearby 
galaxies, along the lines followed by Giuricin et al. (1993) on the basis 
of the NBG (Tully 1988) sample.

\end{document}